\def\selectedoptions{final}

\documentclass[
   \selectedoptions
  ]
  {aipproc}

\newcommand\etpi{\eta\pi^-}
\newcommand\etppi{\eta^{\prime}\pi^-}
\newcommand\etprim{\eta^{\prime}}
\newcommand\ompmpn{\omega\pi^-\pi^0}
\newcommand\bpi{b_1\pi}
\newcommand\jpc{J^{PC}}
\newcommand\omp{1^{-+}}
\newcommand\tripi{\pi^+\pi^-\pi^0}
\newcommand\etdipi{\eta\pi^+\pi^-}
\newcommand\fivpi{\pi^+2\pi^-2\pi^0}
\newcommand\ato{a_2(1320)}
\newcommand\metpi{M_{\eta\pi^-}}
\newcommand\metppi{M_{\eta^{\prime}\pi^-}}
\newcommand\mompmpn{M_{\omega\pi^-\pi^0}}

\def\selectedlayoutstyle{6x9}
\layoutstyle\selectedlayoutstyle

\SetInternalRegister\hbadness{8000} 

\newcommand\doingARLO[2][]{%
  \ifx\mmref\undefined #1\else #2\fi
}

\begin{document}

\title 
      [The $\jpc=\omp$ hunting season at VES]
      {The $\jpc=\omp$ hunting season at VES}


\author{Valery Dorofeev (for VES Collaboration
       \footnote{Amelin~D.V., Dorofeev~V.A., Dzhelyadin~R.I., Gouz~Yu.P., %
    Kachaev~I.A., Karyukhin~A.N., Khokhlov~Yu.A., Ko\-no\-plyan\-nikov~A.K., %
    Konstantinov~V.F., Kopikov~S.V., Kostyukhin~V.V., Matveev~V.D., %
    Nikolaenko~V.I., Ostankov~A.P., Polyakov~B.F., Ryabchikov~D.I., %
    Solodkov~A.A., Solovianov~O.V., Zaitsev~A.M.}~)
}
{  address={Department of Hadron Physics, IHEP, Protvino, Russia, 142284},
  email={dorofeev@mx.ihep.su}}

\copyrightyear  {2001}

\begin{abstract}
We present preliminary results of study of the $\etpi$, $\etppi$ and 
$\bpi$-systems produced in the $\pi^- Be$-interaction at $28~GeV/c$. 
$\jpc=2^{++}$ and $\omp$-waves resulted from the PWA have been fitted in 
each system separately to establish the nature of the $\omp$-wave.
A hypothesis of the $\omp$-wave resonant nature 
in the $\etpi$ and $\etppi$ has no statistically significant preference over 
the non-resonant one. The $\bpi$-system analysis confirms the results of
the $37~GeV/c$-beam data analysis in favor of the resonant treatment of the
bump at $1.6GeV$.   
\end{abstract}

\date{\today}

\maketitle

\section{Introduction}

At present several groups have evidence for $\omp$ meson production, which is 
forbidden for ordinary quarkonia and hence might be a good candidate for a
hybrid \cite{Isgur:Paton:PRD31}, \cite{Close:Page:NPB443}. 
The $\pi_1(1400)$-meson shows up in the $\eta\pi$ final state
in two experiments \cite{Chung:PRD60},\cite{Abele:PLB423},\cite{Abele:PLB446}.
VES group found that the results of the $\etppi$ and $b_1\pi$-system 
Partial-Wave Analysis(PWA) at $37 GeV/c$ agree with the production of the 
higher mass $\pi_1(1600)$ \cite{Khokhlov:NPA663}, evidence for which in 
the $\etppi$ has been also confirmed by E852 at BNL \cite{Ivanov:PRL85}.  
A broad bump observed by VES in the $\omp$-wave intensities of the
$\etprim \pi$, $\bpi$, $\rho\pi$ at $1.6GeV$ was assumed to be a 
single object. The resonant nature results from the analysis of
the $\jpc=2^{++}$ and $\omp$-waves in the $\ompmpn$ \cite{Dorofeev:Frascati}.


A data sample was collected in the 1996 data run at VES spectrometer 
exposed by the $28GeV/c$ momentum $\pi^-$-meson beam.
A detailed description of setup can be found elsewhere \cite{VES:ZPC54}.
Over $5 \times 10^8$ triggers were recorded during the data taking period,
which is $\sim2.5$ more than at $37 GeV/c$.
A trigger is an event with at least two charged tracks from the beam 
interaction moving in the beam direction and falling into the spectrometer 
aperture.
Events of the studied final states production form
subsets of reaction
$\pi^- Be \to \pi^+2\pi^- + k\pi^0(\eta) + Be$, where $k=1,2$. 
To select events of this reaction the following criteria were 
applied to the reconstructed events:
\begin{itemize}
\item 3 reconstructed tracks should form the $\pi^+ \pi^- \pi^-$-system;
\item an interaction vertex must be inside the target;
\item we require for 2-4 $\gamma$-clusters in the calorimeter;
\item a $2 \gamma$-system with mass closest to the mean 
$\pi^0(\eta)$-meson mass within $30(78)MeV$ is identified as 
the $\pi^0(\eta)$-meson.
A 1C-fit to the $\pi^0(\eta)$ mass is applied to the identified
$\gamma\gamma$ pairs;
\item total visible energy $E_{tot}$ must satisfy
the following in-equation: $25GeV<E_{tot}<30GeV$.
\end{itemize} 


\section{Study of the $\etpi$-system}
\begin{figure}[ht]
\includegraphics[width=\textwidth]{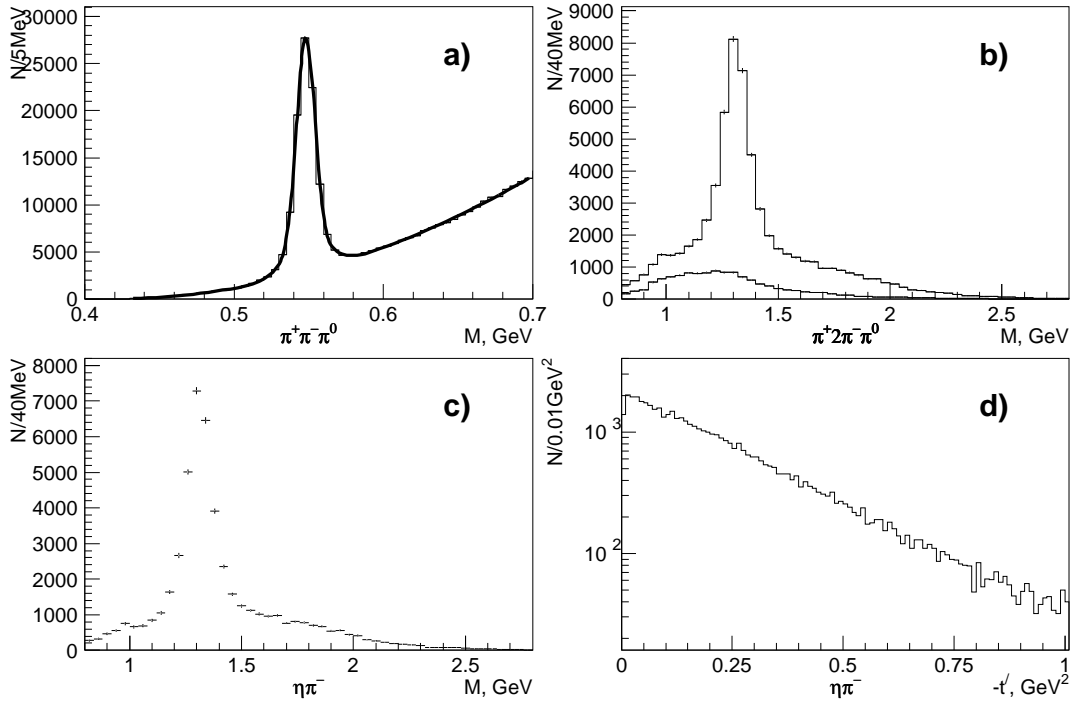}
\caption{Effective mass of the $\tripi$ a), $\pi^+2\pi^-\pi^0$ b) and
 $\etpi$ c). $-t^\prime$ distribution d).} 
\label{fig:1}
\end{figure}

A decay chain $\eta \to \tripi$, $\pi^0\to 2\gamma$ was chosen for  
the $\etpi$ production study.
A clear peak of the $\eta$-meson corresponding to the $\etpi$ 
production is observed in an invariant mass spectrum of the 
$\tripi$-system in fig.~\ref{fig:1}a. The shape of the peak is parametrized
with a sum of two Gaussian functions and resolution of the narrow one 
is equal to $6MeV$.
Shown in fig.~\ref{fig:1}b is the  
invariant mass distribution of the $\pi^+2\pi^-\pi^0$ for events 
from the $\eta$ 
region ($530 <M_{\tripi}< 566MeV$) with superimposed plot for 
sidebands ($502 <M_{\tripi}< 520MeV$ and 
$576 <M_{\tripi}< 594MeV$) which are used for the background estimation.
The background subtracted $\etpi$ invariant mass spectrum
(fig.~\ref{fig:1}c) shows a dominant $\ato$-meson production
with a small bump near $\metpi=1GeV$ of a possible $a_0(980)$-meson 
production.   

A negative value of the invariant four-momentum-transfer between the beam 
and the $\etpi$ squared $-t^\prime$-distribution in fig.~\ref{fig:1}d 
shows the characteristic features of the projective wave production.
Here $t^\prime=t-t_{min}$, where $t_{min}$- is a minimum transfer squared
for a given $\etpi$ mass. 


\subsection{The $\etpi$-system PWA }

We applied the PWA procedure to expand the data in terms of the
partial waves.
An event is considered in the spirit of the isobar model \cite{Herndon:PRD11}
as a production process followed by a chain of the subsequent decays into the 
$\etpi$ and $\eta \to \tripi$.
The production process of the $\etpi$ is assumed to be described by a 
rank one density matrix for each naturality \cite{Hansen:NPB81}. 
An event probability for each naturality  is equal to an average over 
possible non-interfering $\pi^+\pi^-_{i}\pi^0$ combinations $N_{comb}$ of 
a product:
$P_{ev}=\frac{1}{N_{comb}}\sum_{i}^{}{P_i D_i(m_{3\pi}) G_i(m_{3\pi})}$
due to presence of two $\pi^-$'s,
where: $D_i$ - is a probability of the $\eta \to \pi^+\pi^-_{i}\pi^0$ 
decay which corresponds to the decay Dalitz-plot distribution, $G_i$ - is the 
$\eta$-meson shape parameterization being used to 
take into account the $\pi^+\pi^-_{i}\pi^0$ mass resolution.
Here  $P_i$ is an expansion squared of the product of the two pseudo-scalar 
production and decay amplitudes in terms of the partial wave decay amplitudes.
The decay amplitudes are defined by the following set of quantum numbers
$J^PM^{\eta}$, where $J$ stands for a total angular momentum and is equal to
an orbital momentum $L$ in the system of two pseudo-scalars. With
$P$ we denote parity, $M$ - an absolute value of the $J_z$ projection.
Here $\eta$ is exchange naturality. 
The decay amplitude is parametrized with
spherical harmonics \cite{PDG:2000} multiplied by a breakup momentum 
involved in the $L$-th degree.
Under the assumption about the rank of the density matrix the
production amplitudes are denoted as:
$L_0$ for the $M=0$ and $L_\eta$ for the $M=1$ waves. 

The complex production amplitudes are determined from the
extended maximum likelihood fit \cite{Orear}.

A wave set includes 3 subsets of partial waves non-interfering with 
each other. The first is composed of waves with negative
naturality, the second - with positive one and a subset made of a
single wave FLAT which describes the background in the 
$\pi^+2\pi^-\pi^0$-system.

The ambiguous solutions \cite{amb:sol} were found by repeating 
100 times the fit from the random starting values of the fit parameters.



50300 events in the range $0.8<M(\pi^+2\pi^-\pi^0)<2.4GeV$,  
$|t^{\prime}|<1GeV^2$,  with the mass of at least a single $\tripi$
combination in the $\eta$ region or the sidebands were subjected to the PWA.
The PWA has been carried out independently in each of the $40MeV$ bins.
To make the data set cleaner the following additional cuts were applied:
$\cos\Theta_{\pi^0}^{hel}<0.8$ and the track angular separation in projections
$\Delta AX(AY)>10(6)mrad$, where $\Theta_{\pi^0}^{hel}$ is the $\pi^0$
polar angle in the overall CM helicity reference frame.  

\begin{figure}[htb]
\includegraphics[width=\textwidth]{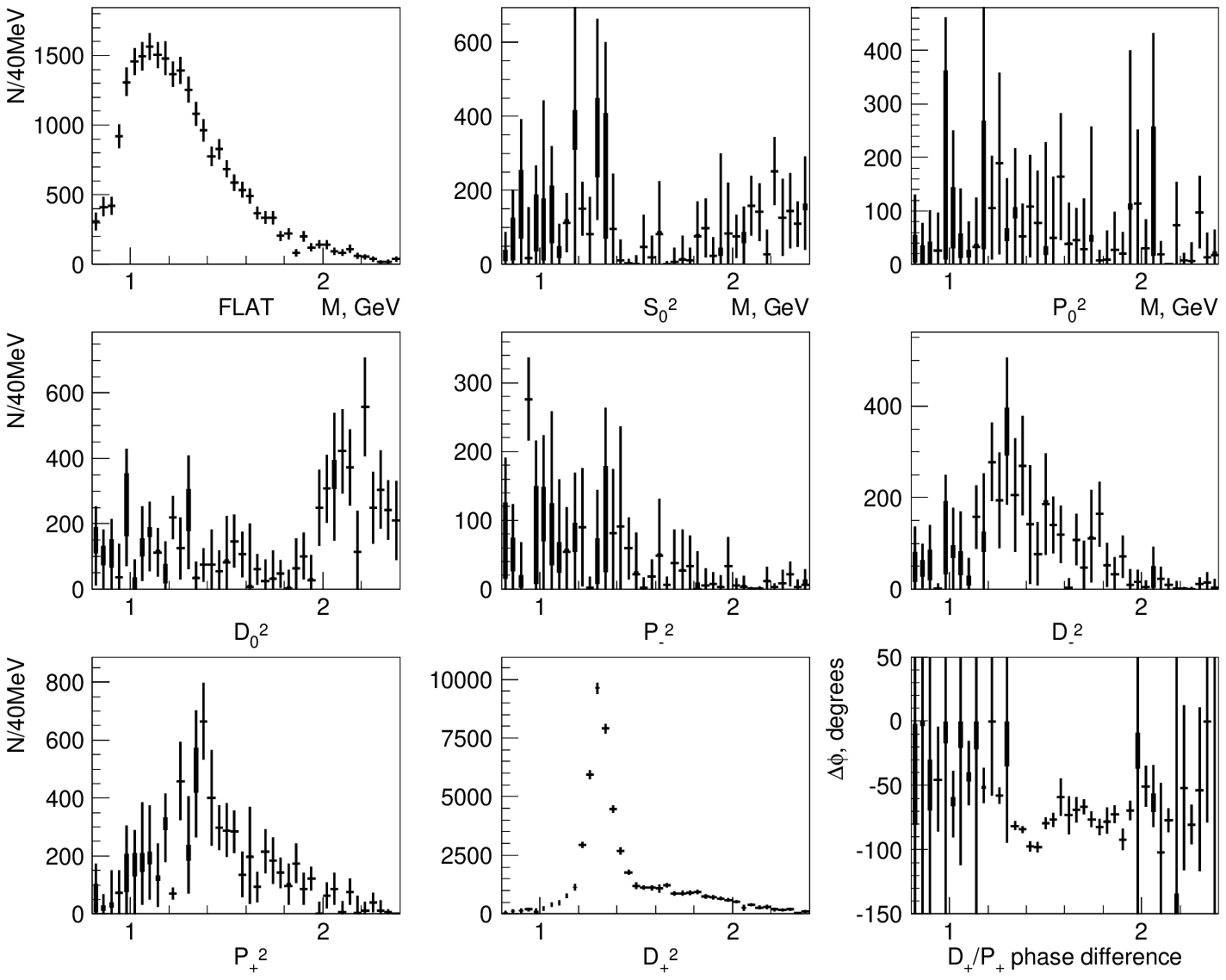}
\caption{The $\etpi$ wave intensities and the $D_+/P_+$ relative phase.}
\label{fig:2}
\end{figure}

In fig.~\ref{fig:2} are shown predicted by the PWA fit acceptance-corrected
wave intensities and a relative phase between the $P_+$ and $D_+$
waves as a function of the $\etpi$ mass. With a thick line is
shown a range of possible ambiguous solutions and with a thin one the maximum
extent of errors.

The intensities of the $S$ and $P$ waves with the negative naturality 
are small and consistent with zero in all the $\metpi$ range. 
The $D_0$ wave has a non-zero contribution in the $\metpi>2GeV$. 
We found a significant $D_-$ intensity in the whole mass
range of unclear nature.

The $D_+$-wave is the dominantly produced wave in the $\etpi$, where
a huge peak corresponding to the $\ato$ production is observed.  
A broad bump with a spike at $\metpi=1.4GeV$ is observed in the $P_+$. 
The spike might results from the fit quality worsening in the $\ato$ region
 and where one should notice the increase of errors.
A sign ambiguous relative phase shows a rapid motion in the $\ato$ 
region.

We have tried also to include the waves with the orbital momentum higher 
than 2 and the waves with $M>1$, but they were found to be not significant.  
The quality of the fit is controlled by comparison of distributions
of variables which describe the $\etpi$-system for the experimental 
and Monte-Carlo data. The Monte-Carlo events were generated according to the
described above model with the values of the production amplitudes
resulted from the best fit solution in a mass bin.  


\subsection{The $\etpi$-system mass-dependent fit}

In order to establish the nature of the $P_+$-wave we have carried out 
a combined mass-dependent(MD) fit of the expected production amplitudes
to the $P_+$ and $D_+$ resulted from the PWA in the whole mass range.
Fitted with the $\chi^2$ method are the $D_+$ and $P_+$ 
intensities and the real and imaginary parts of the production of both 
amplitudes with the corresponding errors in each mass bin.

We assume two general production manners of a resonant and
a non-resonant type. 
The resonant amplitude is parametrized with a Breit-Wigner form:
\begin{displaymath}
A_R=\frac{m_0 \sqrt{\Gamma_0\Gamma_f}}{m_0^2-m^2-i m_0 \Gamma_{tot}},
\hspace{1cm} \Gamma_f=BR_f\Gamma_0\frac{m_0}{m}\frac{q}{q_0}
\left [ \frac{B_L(q)}{B_L(q_0)} \right ]^2
\end{displaymath}
where: $m$ - the $\eta\pi$ mass, $m_0$ - a resonance mass,
$\Gamma_0$ - a nominal width, $\Gamma_f$ - a partial width in the final 
channel, $\Gamma_{tot}$ - a total width,
$q(q_0)$ - a breakup momentum of the $\eta\pi$ with the $m(m_0)$ mass,
$B_L(q)$ - a barrier factor \cite{Hippel:PRD5},
$BR_f$ - a branching ratio into the $f$-th final state.

By the notion background we just mean the non-resonant production manner
without going into the nature of the process. The main difference of the
background type from the resonant one is absence of phase motion.  
The background amplitude is parametrized with the following free form
$A_B=LIPS_l \times (m-m_t)^{\alpha}\exp{(-\beta (m-m_t))}$,
where: $LIPS_l(m)$ - a phase space, $m_t$ - a threshold mass,
$\alpha$ and $\beta$ are shape parameters.

A procedure of the wave construction from the objects is in the following.
At first a wave is formed of evident structures such as the 
$\ato$ in the $D_+$-wave. If there are no such objects we take 
an object of any production type arbitrarily. After doing the fit to thus
constructed wave we then necessarily try to describe the data with
the object of the alternative type.  
At the next steps we add consistently another objects of both types to
both waves until we will succeed in the description of the wave 
intensities and their interference in the whole mass region. 
All objects are added coherently with their own mass-independent phase.
The parameters of the MD fit are:
the mass, width and the number of the resonance events, 
the shape parameters and the number of the background events,
the phase relative to a reference object.
  
\begin{figure}[htb]
\includegraphics[width=\textwidth,bb=0 100 424 340]{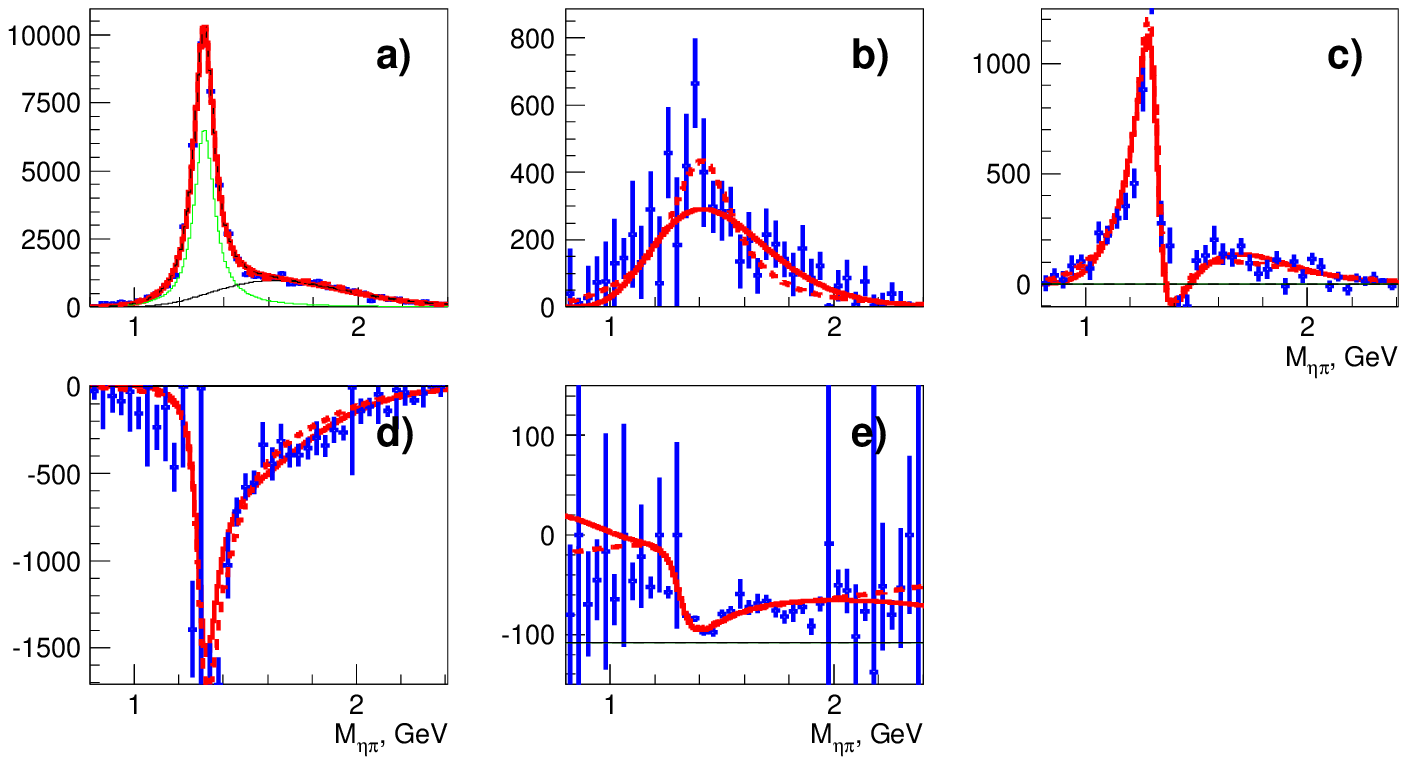}
\caption{The $\etpi$ $D_+$ a) and $P_+$ b) intensities. The real c) and 
imaginary d) parts 
of $D_+P_+$ product. The $D_+/P_+$ relative phase e).
The smooth curves are the MD fit results.}
\label{fig:3}
\end{figure}

\begin{center}
\begin{table}[htb]
\begin{tabular}{c|c|c|c|c} \hline
$D_+$ bkg & $a_2^{\prime}$ & $P_+$ bkg & $\pi_1$& $\chi^2/NDF$ \\ \hline
    +     &                &     +     &        &  244/149     \\
          &      +         &     +     &        &  422/149     \\
    +     &                &           &   +    &  224/149     \\
          &      +         &           &   +    &  219/149     \\
    +     &                &     +     &   +    &  176/145     \\
    +     &                &     +     & E852   &  187/147     \\ \hline
\end{tabular}
\caption{Comparison of the MD fit variants}
\label{tab:2}
\end{table}
\end{center}
To begin with we have fitted the $\ato$ Breit-Wigner to the $D_+$ intensity 
and failed to describe the $D_+$ wave at masses $>1.5GeV$.
Then we have applied the described above procedure.
At first the $D_+$ was fitted to a sum of the $\ato$ and the background 
and the $P_+$ to the background. The second fit was carried out 
with the $D_+$ wave background replaced by the resonance,
which will be called further as an $a_2^{\prime}$-meson for convenience.
There have just been reported the evidence for the $a_2(1750)$ \cite{Belle}. 
The next steps were to fit the $P_+$ wave to
the resonance, called a $\pi_1(1400)$. The evidence for it have been reported 
by E852 and CBAR groups \cite{Chung:PRD60}, \cite{Abele:PLB423},
\cite{Abele:PLB446} and which production did not contradict to VES 
results \cite{Zaitsev:Hadron97}.
The results of the MD fits in the case, when the $D_+$ has been fitted to
a sum of the $\ato$ and the background and the $P_+$ to the 
background or to the $\pi_1$ are shown in fig.~\ref{fig:3} 
with thick solid and dashed curves respectively. 

The summary of the MD fit results are presented in tab.~\ref{tab:2}.
The objects which make up the fitted partial waves are marked by 
the plus sign in a row.  A $\chi^2$ over the number of degrees of freedom of 
the corresponding MD fit is shown in the fifth column.

From the comparison of the first four rows one can conclude that there are no 
statistically significant difference between a single resonant and non-resonant
object description of the $P_+$. 

The fits of the $P_+$ to a sum of the $\pi_1$ and the 
background are shown in the last two rows. A fit with free $\pi_1$ parameters
resulted in the following values: $M_{\ato} = 1316 \pm 1MeV $, 
$\Gamma_{\ato} = 113 \pm 3MeV$, and $M_{\pi_1} = 1316 \pm 12MeV$, 
$\Gamma_{\pi_1} = 287 \pm 25MeV$. The quoted errors are statistical only.  
Statistical significance is nearly the same as for the fit with the 
$\pi_1$ parameters fixed to the values reported by E852
\cite{Chung:PRD60}.
A closeness of the $\pi_1$ mass to the $\ato$ mass may indicate that there is
an underestimated influence of the $\ato$ on the $P_+$.


\section{Study of the $\etppi$-system }

\begin{figure}[htb]
\includegraphics[width=\textwidth,bb=0 100 424 340]{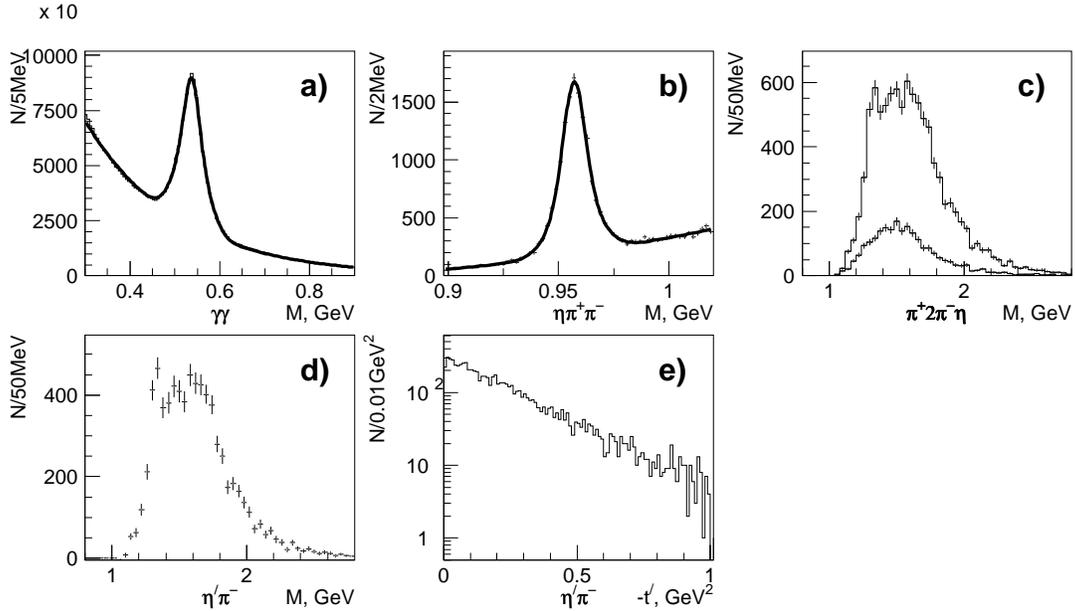}
\caption{Effective mass of the $\gamma\gamma$ a), $\etdipi$ b), 
$\eta\pi^+2\pi^-$ c) and
 $\etppi$ d). $-t^\prime$ distribution e).} 
\label{fig:5}
\end{figure}

A decay chain $\etprim \to \etdipi $, $\eta\to 2\gamma$ was
chosen for the $\etppi$ production study.
A clear peak of the $\etprim$ corresponding to the $\etppi$ 
production is observed in the invariant mass spectrum of the 
$\etdipi$ in fig.~\ref{fig:5}b. 
Shown in fig.~\ref{fig:5}c is the invariant mass 
distribution of the $\eta\pi^+2\pi^-$ for events from $\etprim$ 
region ($940 <M_{\etdipi}< 976MeV$) with superimposed plot for the 
$\etprim$ sidebands ($912 <M_{\etdipi}< 930MeV$ and 
$986 <M_{\etdipi}< 1004MeV$) which are used for the background estimation.
The background subtracted $\etppi$ invariant mass spectrum
(fig.~\ref{fig:5}d) shows a clear $\ato$ peak and a broad large
bump at $\metpi=1.6GeV$, which strongly differs from the corresponding 
distribution of the $\etpi$.   

A $-t^{\prime}_{\etppi}$ distribution in fig.~\ref{fig:5}e shows 
characteristic features of the projective wave production like for the $\etpi$.



10700 events in the range $1.15<M(\eta\pi^+2\pi^-)<2.15GeV$,  
$|t^{\prime}|<1GeV^2$,  with the mass of at least a single $\etdipi$
combination in the $\etprim$-meson region or the sidebands were subjected to 
the PWA. The PWA has been carried out independently in each of the $50MeV$ 
wide bins.

\begin{figure}[htb]
\includegraphics[width=\textwidth]{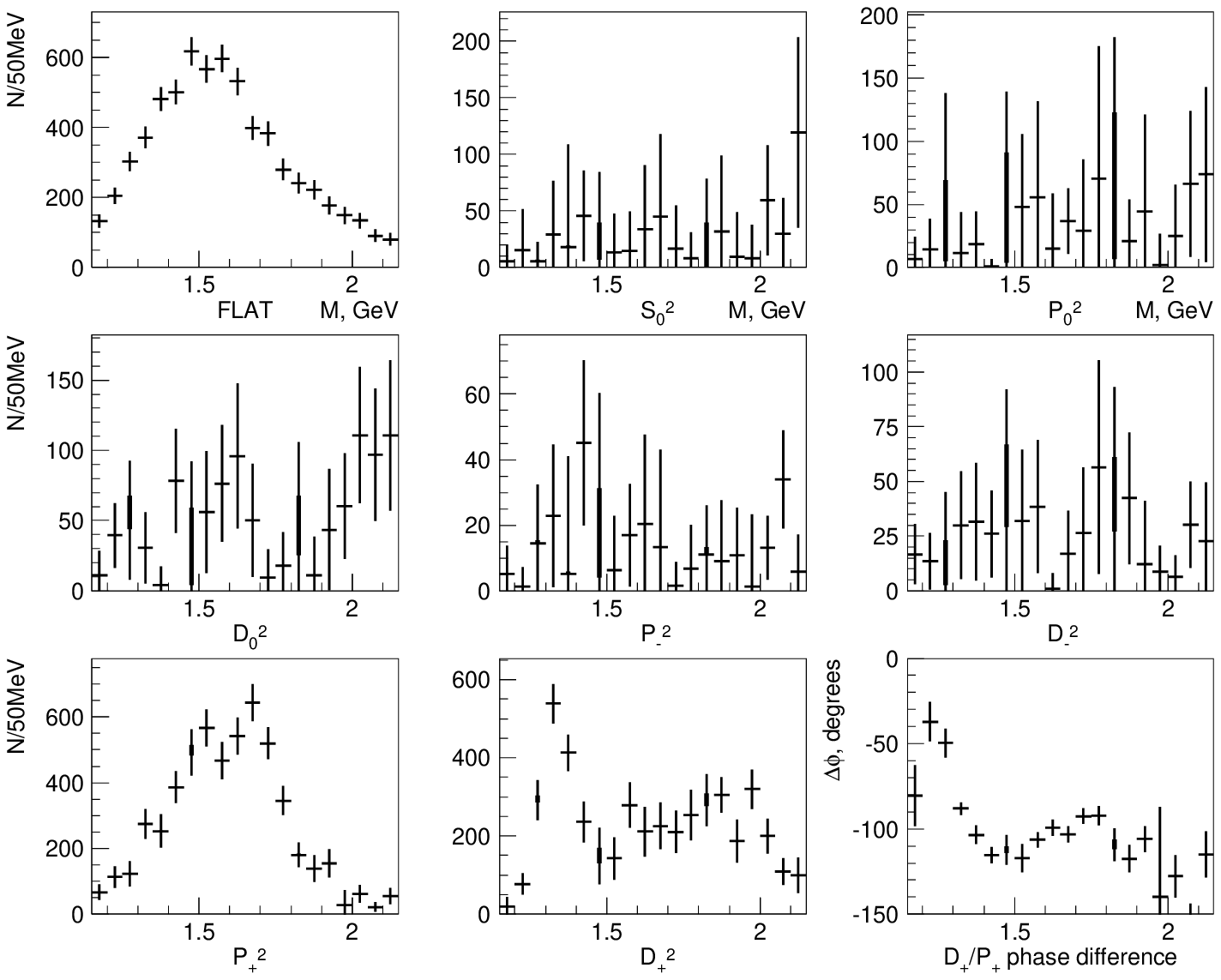}
\caption{The $\etppi$ wave intensities and the $D_+/P_+$ relative phase.}
\label{fig:6}
\end{figure}

We have applied the same procedure as for the $\etpi$-system PWA, described
in the appropriate section.
In fig.~\ref{fig:6} are shown predicted by the PWA fit acceptance-corrected
wave intensities and a phase difference of the $P_+$ and $D_+$
as a function of the $\etppi$ mass. With a thick line is
shown a range of possible ambiguous solutions and with a thin one the 
maximum extent of errors.

The intensities of the negative naturality waves are small and nearly 
consistent with zero in all the $\metppi$ range. The $D_0$ wave only shows
some non-negligible contribution in the regions $1.4 \div 1.6GeV$ and
$>1.9GeV$. 

The $\ato$ peak is observed in the $D_+$ intensity. 
But contrary to the $\etpi$-system behavior this wave shows a broad structure 
in the region $\metppi>1.5GeV$. The rate of this high mass production 
relative to the $\ato$ is larger for the $\etppi$ than for $\etpi$.

A broad bump of a $\sim 350MeV$ width is observed in the $P_+$ at 
$\metppi=1.6GeV$. A sign ambiguous relative phase shows a rapid motion in 
the $\ato$ mass region.

We have tried also to include the waves with the orbital momentum higher 
than 2 but they were found to be not significant.  
The quality of the fit is controlled by comparison of distributions
of variables which describe the $\etppi$-system for the experimental 
and Monte-Carlo data.   


\begin{figure}[htb]
\includegraphics[width=\textwidth,bb=0 100 424 340]{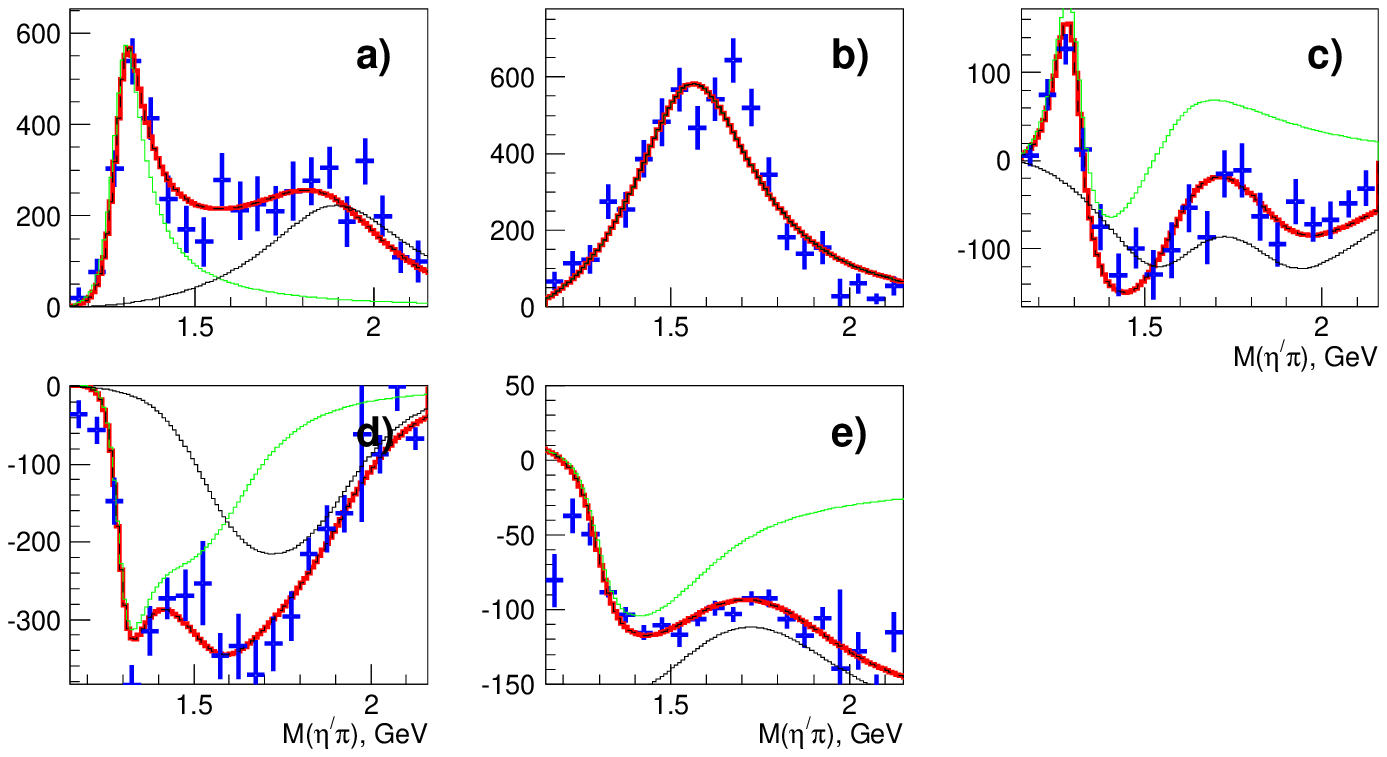}
\caption{The $\etppi$ $D_+$ a) and $P_+$ b) intensities. The real c) and 
imaginary d) parts of $D_+P_+$ product. The $D_+/P_+$ relative phase e). 
The smooth curves are the MD fit results.}
\label{fig:7}
\end{figure}

To establish the nature of the $P_+$-wave we have carried out the 
MD fit following the procedure which had been applied for the $\etpi$ study. 
The $D_+$ partial wave amplitude was parametrized with a sum
of the $\ato$ Breit-Wigner function and the background or an $a_2^{\prime}$. 
This additional object was included to describe the broad
bump in the region $\metppi>1.5GeV$. 
The $P_+$ partial wave was parametrized with the background or with the 
$\pi_1(1600)$. 
The indication to the possible resonance nature of the bump at $1.6GeV$
was reported by our group as a result of the $\etppi$ PWA at $37GeV/c$
\cite{Khokhlov:NPA663}.
Later the observation of the resonance have been reported by E852
\cite{Ivanov:PRL85}. 
The result of the MD fit in the case, when the $D_+$ is fitted to
a sum of the $\ato$ and the $a_2^{\prime}$  and the $P_+$ to the 
resonance only is shown in fig.~\ref{fig:7}. 

We have found no statistically significant preference of the fit with the 
$P_+$ saturated with the resonant over the non-resonant description.
In the case of the $P_+$ partial wave amplitude parameterization
with a sum of the resonance and the background the statistical significance
is larger than in the previous cases. However the $\ato$ width
was found to be somewhat larger than it could be expected from the 
resolution study. In the cases, when the $D_+$ is composed of
the $\ato$ and a $a_2^{\prime}$, while the $P_+$ includes
the $\pi_1$, the $a_2^{\prime}$ mass was found to be somewhat
$1.8 \div 1.9 GeV$ and the width $0.4 \div 0.5 GeV$, while 
one might expect to find the same parameters as for the 
$a_2^{\prime}(1750)$. 
   	 
The above consideration of results prevents us from drawing the unambiguous 
conclusion about the nature of the $P_+$-wave in the $\etppi$.


\section{Study of the $\ompmpn$-system }
\begin{figure}[htb]
\includegraphics[width=\textwidth,bb=0 100 424 340]{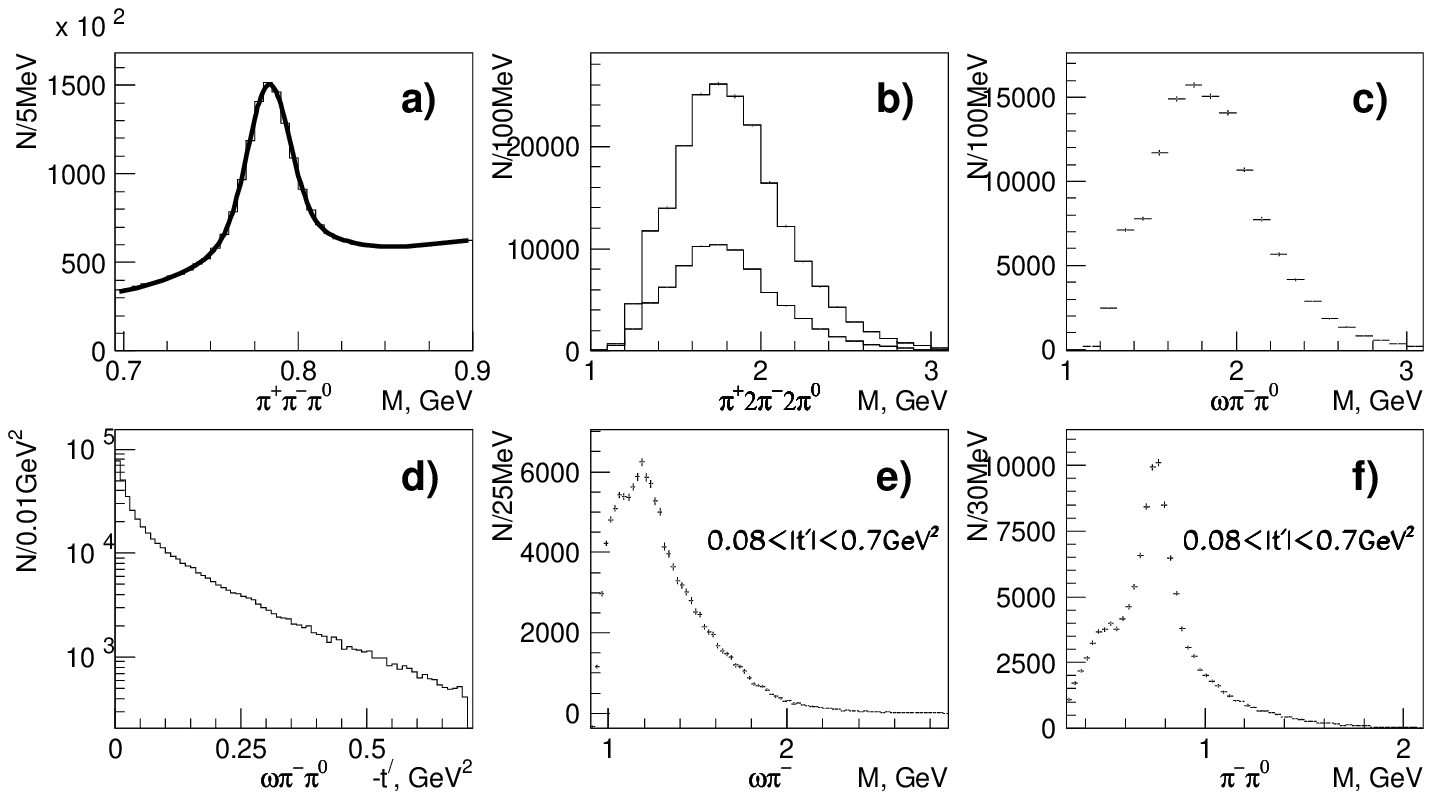}
\caption{Effective mass of the $\tripi$ a), $\pi^+2\pi^-2\pi^0$ b) and
 $\ompmpn$ c). $-t^\prime$ distribution d). Effective mass of
the $\omega\rho^-$ e) and $\pi^-\pi^0$ f) from the $\ompmpn$
for $0.08<|t^{\prime}|<0.7GeV^2$.}
\label{fig:8}
\end{figure}

A decay mode $\omega \to \tripi$ 
was chosen for the $\ompmpn$ production study.
A clear peak of the $\omega$ meson corresponding to the $\ompmpn$ 
production is observed in the invariant mass spectrum of the 
$\tripi$ in fig.~\ref{fig:8}a. 
Shown in fig.~\ref{fig:8}b is the invariant mass 
distribution of the $\fivpi$ for events from the $\omega$ 
region ($758 <M_{\tripi}< 808MeV$) with superimposed plot for the 
$\omega$ sidebands ($684 <M_{\tripi}< 719MeV$ and 
$846 <M_{\tripi}< 871MeV$) which are used for the background estimation.
The background subtracted $\ompmpn$ invariant mass spectrum
(fig.~\ref{fig:8}c) shows an $\ato$ shoulder 
and a broad large bump at $\mompmpn=1.8GeV$.   

A $-t^{\prime}_{\ompmpn}$-distribution 
in fig.~\ref{fig:8}d shows a sharp peak in the low $|t^{\prime}|<0.08GeV^2$ 
(LT) region, corresponding to the coherent production on a nucleus,
which flattens in the high $|t^{\prime}|$ region 
($0.08<|t^{\prime}|<0.7GeV^2$) and which will be called further the HT. 
This behavior in the HT-region 
corresponds to the incoherent production on a nucleon. The projective wave
production dominates in the HT-region. Therefore we have divided a data 
sample into the LT- and HT-sets, which were analyzed separately.     
A large $\rho^-$-meson peak is observed in the
invariant mass spectrum of the $\pi^-\pi^0$, produced with the
$\omega$(see fig.~\ref{fig:8}f). 
This means the dominant production of the $\omega\rho^-$.
There is a peak of the $b_1(1235)^-$-meson in the $\omega\pi^-$ invariant 
mass spectrum shown in fig.~\ref{fig:8}e, corresponding to the 
$b_1(1235)\pi$ production. The same is for the $b_1(1235)^0$. 

\begin{figure}[htb]
\includegraphics[width=\textwidth,bb=0 100 424 340]{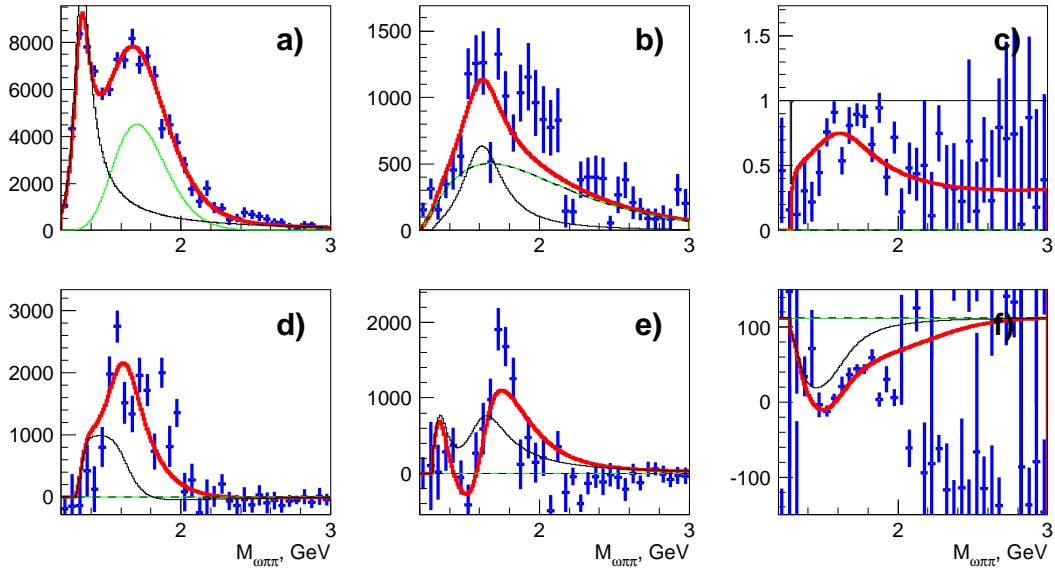}
\caption{The $2^+1^+S2(\omega\rho)$ a) and $1^-1^+S1(b_1\pi)$ b) intensities.
A coherence parameter c) \cite{Hansen:NPB81}.
The real d) and imaginary e) parts of their non-diagonal $\rho$-matrix element.
The $1^-$ phase relative to $2^+$ f). 
The smooth curves are the MD fit results.}
\label{fig:9}
\end{figure}

A PWA of 
284000 events in the range $1.2<M(\pi^+2\pi^-2\pi^0)<3GeV$, 
$0.08<|t^{\prime}|<0.7GeV^2$, $\max_{\pi^{\pm}}(\cos\Theta^{hel})<0.92$  
with the mass of at least one $\tripi$-combination in the 
$\omega$ region or in the sidebands has been carried out.
Here $\Theta^{hel}$ is a track polar angle in the overall helicity CM 
reference frame.
Events in each of the $50MeV$ or $100MeV$ wide bins were fitted separately.
The Illinois method \cite{Hansen:NPB81} was used in the PWA, however with 
an important improvement. We added the possibility to restrict a rank 
of the density matrix \cite{Chung:PRD11}.
The partial waves are denoted by the $J^PM^{\eta}LS(isobar-bachelor)$, 
where $S$ stands for a total spin in the isobar-bachelor system. 
The partial waves 
of the $\omega\rho^-$, $b_1(1235)\pi$, $\rho_3(1690)\pi$, $\rho_1(1450)\pi$
are included in the wave set. The results of the PWA are consistent with
our previous analysis of the $\ompmpn$ at $37GeV/c$ \cite{Amelin:PAN62}.
A $2^+1^+S2(\omega\rho)$ shown in fig.~\ref{fig:9}a
was found to be a dominant wave  with a clear
$\ato$ peak and a broad bump at $\mompmpn=1.7GeV$. 
A significant $1^-1^+S1(b_1\pi)$ wave shown in fig.~\ref{fig:9}b
is observed with a broad bump at 
$\mompmpn=1.6GeV$, which is in the maximum $\sim15\%$ of the $\ato$ 
peak height. 
The important feature of the combined $2^+$ and $1^-$ wave behavior 
demonstrated in fig.~\ref{fig:9}c is a region bounded to  
$1.5 < \mompmpn < 2GeV$, where the coherence is significantly 
non-zero. It should be noticed in fig.~\ref{fig:9}f
an $80^{\circ}$ rise of the $1^-$-wave phase relative to the $2^+$-wave phase
right in this region and which may be attributed to a $1^-$ resonance.   

The MD fit has been carried out to establish the origin of the $1^-1^+$-wave
in the $b_1\pi$. The high intensity $2^+1^+S2(\omega\rho)$ was selected 
to be a reference wave. 
The fit has been done in the same way like the $\etpi$ fit.
However absence of unambiguous knowledge about the nature of the bump at 
$\mompmpn=1.7GeV$ and an extension of the density matrix rank to three in 
the PWA requires inclusion of additional fitting parameters in the MD fit.
The na{\"i}ve model is to describe the $2^+$ wave by a sum of the $\ato$ 
Breit-Wigner and something, which has a form of the bump at $1.7GeV$,
while the $1^-$ may be constructed from a broad incoherent background and
from a resonance, which is coherent to the $2^+$ bump.  

The fit describes the interference pattern satisfactorily
as seen in fig.~\ref{fig:9}. 
We have tried another hypotheses: to fit the $1^-$ wave to a partially 
coherent background or to make up the $2^+$ wave from a
$a^{\prime}_2$ in addition to the $\ato$.
However we failed to find qualitatively better solution.
The data are consistent with the resonant description of the 
$1^-1^+S1(b_1\pi)$ with the mass $1.6GeV$ and the width $0.33GeV$.


\section{ Conclusions}

We have carried out the PWA of the $\eta\pi^-$, $\eta^{\prime}\pi^-$, 
$b_1\pi$ systems produced in $\pi^-Be$ interaction at $28GeV/c$. 

We have tried to understand the nature of the observed partial waves 
in each system by the fit to the resonant and non-resonant description.  

\begin{itemize}
\item We observed the $P_+$ in the $\eta\pi^-$ which exhibits 
the phase motion with respect to the $D_+$ in the coherent model.
We could not unambiguously establish the nature of the $P_+$ in
the $\eta\pi^-$.   
\item The production of the $\eta^{\prime}\pi^-$ is dominated 
by the $P_+$ and $D_+$. The common analysis of the 
observed $\eta^{\prime}\pi^-$ partial waves gives no preference to 
the hypothesis of the $\pi_1(1600)$ production.  
\item We confirmed the results of the $b_1\pi$ analysis of the
$37GeV/c$-beam data. The broad bump 
has been observed in the $1^-1^+S1(b_1\pi)$ of the $\ompmpn$. 
The simultaneous fit with the $2^+1^+S2(\omega\rho)$ favors
the existence of a resonance in the $b_1\pi$ with the mass $\sim 1.6 GeV$
and the width $\sim 330 MeV$.  
\item The PWA of the $37GeV/c$-beam data sample in the several $t^{\prime}$
intervals shows a wide bump in the 
$J^PM^{\eta}(isobar-bachelor)=1^-1^+(\rho\pi)$-wave 
at $M_{\rho\pi}<2GeV$ \cite{Kachaev}.
\end{itemize}


\begin{theacknowledgments}
This is supported, in part, by INTAS-RFBR 97-02-71017, INTAS-RFBR 00-02-16555,
RFBR 00-15-96689 grants.
\end{theacknowledgments}


\doingARLO[\bibliographystyle{aipproc}]
          {\ifthenelse{\equal{\AIPcitestyleselect}{num}}
             {\bibliographystyle{arlonum}}
             {\bibliographystyle{arlobib}}
          }
\bibliography{hadron2001_rep2}

\hyphenation{Post-Script Sprin-ger}
\begin{thebibliography}{20}
\expandafter\ifx\csname natexlab\endcsname\relax\def\natexlab#1{#1}\fi
\providecommand{\enquote}[1]{``#1''}
\expandafter\ifx\csname url\endcsname\relax
  \def\url#1{\texttt{#1}}\fi
\expandafter\ifx\csname urlprefix\endcsname\relax\def\urlprefix{URL }\fi

\bibitem[Isgur and Paton(1985)]{Isgur:Paton:PRD31}
Isgur, N., and Paton, J., \emph{Phys. Rev.}, \textbf{D31}, 2910 (1985).

\bibitem[Close and Page(1995)]{Close:Page:NPB443}
Close, F.~E., and Page, P.~R., \emph{Nucl. Phys.}, \textbf{B443}, 233 (1995).

\bibitem[Chung et~al.(1999)]{Chung:PRD60}
Chung, S.~U., et~al., \emph{Phys. Rev.}, \textbf{D60}, 092001 (1999).

\bibitem[Abele et~al.(1998)]{Abele:PLB423}
Abele, A., et~al., \emph{Phys. Lett.}, \textbf{B423}, 175 (1998).

\bibitem[Abele et~al.(1999)]{Abele:PLB446}
Abele, A., et~al., \emph{Phys. Lett.}, \textbf{B446}, 349 (1999).

\bibitem[Khokhlov et~al.(2000)]{Khokhlov:NPA663}
Khokhlov, Y., et~al., \emph{Nucl. Phys.}, \textbf{A663}, 596 (2000).

\bibitem[Ivanov et~al.(2001)]{Ivanov:PRL85}
Ivanov, E.~I., et~al., \emph{Phys. Rev. Lett.}, \textbf{85}, 3977 (2001).

\bibitem[Dorofeev et~al.(1999)]{Dorofeev:Frascati}
Dorofeev, V.~A., et~al., \enquote{New results from VES}, in \emph{Workshop on
  Hadron Spectroscopy}, edited by B.~T., A.~Feliciello, and A.~Filippi,
  Frascati Phys. 15, 1999, p. 999.

\bibitem[Beladidze et~al.(1992)]{VES:ZPC54}
Beladidze, G., et~al., \emph{Zeit. fur Phys.}, \textbf{C54}, 235 (1992).

\bibitem[Herndon et~al.(1975)]{Herndon:PRD11}
Herndon, D.~J., S{\"o}ding, P., and Cashmore, R., \emph{Phys. Rev.},
  \textbf{D11}, 3165 (1975).

\bibitem[Hansen et~al.(1974)]{Hansen:NPB81}
Hansen, J.~D., Jones, G., Otter, G., and Rudolph, G., \emph{Nucl. Phys.},
  \textbf{B81}, 403 (1974).

\bibitem[Groom et~al.(2000)]{PDG:2000}
Groom, D.~E., et~al., \emph{Eur. Phys. J.}, \textbf{C15}, 1--878 (2000).

\bibitem[Orear(1958)]{Orear}
Orear, J., Notes on statistics for physicists (1958), in UCRL-8417.

\bibitem[Sadovsky(1991)]{amb:sol}
Sadovsky, S.~A., On the ambiguities in the partial-wave analysis of $\pi^-p \to
  \eta\pi^0n$ reaction (1991), preprint IHEP-91-75.

\bibitem[von Hippel and Quigg(1972)]{Hippel:PRD5}
von Hippel, F., and Quigg, C., \emph{Phys. Rev.}, \textbf{D5}, 624 (1972).

\bibitem[Hou(2001)]{Belle}
Hou, S., in this proceedings (2001).

\bibitem[Zaitsev et~al.(1997)]{Zaitsev:Hadron97}
Zaitsev, A.~M., et~al., \enquote{Search for exotics in $I^GJ^P=1^-1^-$,
  $1^-0^-$ and $0^+2^+$ waves}, in \emph{CP432, Hadron Spectroscopy: Seventh
  Internaltional Conference}, edited by S.~U. Chung and W.~H. J., 1997, pp.
  461--470.

\bibitem[Chung and Trueman(1975)]{Chung:PRD11}
Chung, S.~U., and Trueman, T.~L., \emph{Phys. Rev.}, \textbf{D11}, 633 (1975).

\bibitem[Amelin et~al.(1999)]{Amelin:PAN62}
Amelin, D.~V., et~al., \emph{Phys. Atom. Nucl.}, \textbf{62}, 445--453 (1999).

\bibitem[Kachaev(2001)]{Kachaev}
Kachaev, I.~A., in this proceedings (2001).

\end{thebibliography}

\end{document}